\documentclass[conference]{IEEEtran}
\IEEEoverridecommandlockouts

\usepackage[T1]{fontenc}
\usepackage{cite}
\usepackage{amsmath,amssymb,amsfonts,mathtools}
\usepackage{algorithmic}
\usepackage{graphicx}
\usepackage{textcomp}
\usepackage{graphicx}
\usepackage{xcolor}
\usepackage{bbm}
\usepackage{caption}
\usepackage{subcaption}

\def\BibTeX{{\rm B\kern-.05em{\sc i\kern-.025em b}\kern-.08em
    T\kern-.1667em\lower.7ex\hbox{E}\kern-.125emX}}
    
\usepackage{tikz,pgfplots}
\pgfplotsset{compat=1.17}
\usetikzlibrary{plotmarks}

\usepackage{soul}
\setstcolor{red} 
\setul{0.3ex}{0.7pt} 
\definecolor{darkgray}{rgb}{0.66, 0.66, 0.66}
\setulcolor{darkgray}
    
\usepackage{cancel}

\def\mathcolor#1#{\@mathcolor{#1}}
\def\@mathcolor#1#2#3{%
  \protect\leavevmode
  \begingroup
    \color#1{#2}#3%
  \endgroup
}

\begin{document}

\title{Soft Syndrome Decoding of Quantum LDPC Codes for Joint Correction of Data and Syndrome Errors\\
\thanks{
This work is generously supported by the National Science Foundation (NSF) under grants
    CCF-2100013, 
    CCF-2106189, 
    CCSS-2027844, 
    CCSS-2052751, 
    ERC-1941583, 
    CIF-1855879, 
and 
    NASA under the SURP Program.
}
}

\author{\IEEEauthorblockN{
Nithin Raveendran, 
Narayanan Rengaswamy, 
Asit Kumar Pradhan,
and Bane Vasi\'{c}}
 \IEEEauthorblockA{\textit{Department of Electrical and Computer Engineering, The University of Arizona, Tucson, AZ, 85721 USA}\\ 
 Email: \{ nithin , narayananr , asitpradhan \}@arizona.edu, vasic@ece.arizona.edu}}

\maketitle

\begin{abstract}
Quantum errors are primarily detected and corrected using the measurement of syndrome information which itself is an unreliable step in practical error correction implementations.
Typically, such faulty or noisy syndrome measurements are modeled as a binary measurement outcome flipped with some probability.
However, the measured syndrome is in fact a discretized value of the continuous voltage or current values obtained in the physical implementation of the syndrome extraction. 
In this paper, we use this ``soft'' or analog information without the conventional discretization step to benefit the iterative decoders for decoding quantum low-density parity-check (QLDPC) codes.
Syndrome-based iterative belief propagation decoders are modified to utilize the syndrome-soft information to successfully correct both data and syndrome errors simultaneously, without repeated measurements.
We demonstrate the advantages of extracting the soft information from the syndrome in our improved decoders, not only in terms of comparison of thresholds and logical error rates for quasi-cyclic lifted-product QLDPC code families, but also for faster convergence of iterative decoders. In particular, the new BP decoder with noisy syndrome performs as good as the standard BP decoder under ideal syndrome. 
\end{abstract}



\section{Introduction}
\label{Sec:Intro}
Quantum error correction (QEC) is a necessary step in realizing scalable and reliable quantum computing. 
For fault-tolerant QEC, quantum low-density parity-check (QLDPC) codes are advantageous over surface codes, the current leading candidate~\cite{mackay_quantum, Hanzo_survey_decoders,gottesman_fault_tolerant_ldpc,quantum_expander_codes,ReviewQLDPC_Breuckmann}, in terms of the scaling of qubit overhead and minimum distance. 
Recent breakthroughs related to various product constructions of QLDPC codes provided constructions of ``good'' QLDPC code families, i.e., codes with finite asymptotic rate and relative minimum distance~\cite{quantum_expander_codes,hastings2020fiber, breuckmann2020balanced,panteleev2021quantumLinearMinDLocalTestable,QuantumTannerCodes}. 
Similarly, there has been significant progress in improving the iterative decoding performance of finite-length QLDPC codes using post-processing and heuristic techniques \cite{Panteleev_Degenerate_BP_OSD2021, refinedBP_QLDPC_2020, Poulin_NN_BP, Rigby2019_ModifiedBP_QLDPC,NithinTrappingSetQLDPC}. 
However, the QLDPC decoding problem still has unanswered questions and, in particular, faster decoders for QLDPC codes are needed to meet the stringent timing constraints in hardware. 
The problem is even more challenging than classical LDPC code decoding since we have to overcome syndrome measurement errors in addition to the errors on physical qubits. 
To address these errors without repeating the measurements, data-syndrome codes and ``single-shot'' error correction codes have been proposed~\cite{ashikhmin_error_in_syndrome,ashikhmin_data_syndrome_codes,bombin_single_shot,campbell2019SingleShottheory}. 
However, noisy syndrome measurement has not been analyzed under iterative decoding, which is of great interest while decoding these good QLDPC codes. 

The first, and primary, challenge is that when a decoder uses a noisy syndrome as input, it significantly and non-trivially alters the iterative decoding dynamics. Secondly, in the absence of a referent ``true'' syndrome, it is difficult to formulate the halting condition, and an iterative decoder can, in principle, run indefinitely without finding the error pattern. A na{\"{\i}}ve binary quantization of the syndrome would lead to an estimated error that is possibly matched to a wrong syndrome.

In the literature, noisy measurements are typically modeled by a simple phenomenological error model in which binary measurement outcomes are flipped with some probability. 
However, the measurement is typically more informative since it corresponds to continuous voltages, currents, or discrete photon counts based on the specific physical implementation. 
Instead of quantizing this analog information to a binary outcome and losing crucial information, one can leverage the soft syndrome in the iterative decoder and modify the update rules accordingly. 
In a recent work by Pattinson \emph{et al.}~\cite{ImprovedDecodingSoftInfo_2021}, syndrome measurement outcomes beyond simple binary values were considered. 
The decoders used for surface codes, such as minimum weight perfect matching and union-find, were modified accordingly to obtain higher decoding thresholds. 

In this work, we model a realistic syndrome measurement by a perfect measurement with an ideal bipolar ($\pm 1$) outcome followed by a noisy channel that leads to a continuous soft outcome. We restrict our attention to the simple setting of a symmetric Gaussian noise on each of the ideal syndrome measurements. 
In the context of iterative decoding, we \emph{modify} the min-sum algorithm (MSA) based decoder (a hardware friendly variant of the belief propagation (BP) decoder) to use these soft syndrome values as additional information to drive the message passing algorithm towards convergence. 
This iterative decoder for QLDPC codes that employs \emph{soft syndrome information} is novel and is one of the main contributions of this paper.
It is also unprecedented in the setting of iterative decoding of classical LDPC codes, although faulty components in decoders have been studied~\cite{one_error_drives_another_tcomm_2015,16_ITA_stoch_resonance_ml,Nithin_stochastic_GallagerB_Quantum}.
Furthermore, even if the code does not have the single-shot property, our strategy allows one to perform effective decoding \emph{without repeating syndrome measurements}, which is attractive practically.

We consider three scenarios for decoding:
\begin{enumerate}

\item \textbf{Perfect Syndrome:} In this case, there is no noise added to the syndrome and we use the standard MSA decoder.

\item \textbf{Hard Syndrome:} Here, there is Gaussian noise added to the syndrome, but this continuous value is thresholded to input a hard syndrome to the standard MSA decoder.

\item \textbf{Soft Syndrome:} Now, as in the hard syndrome case, the syndrome is noisy and it is thresholded to produce a sign ($\pm 1$), but the magnitude is converted into a log-likelihood ratio (LLR) to also provide soft information. The decoder in this case is our new modified MSA, where the check node update rules are modified and, specifically, the LLR corresponding to the syndrome is utilized and also updated during the check node update.

\end{enumerate}

\begin{figure}[t]
    \centering
    \includegraphics[width=0.48\textwidth]{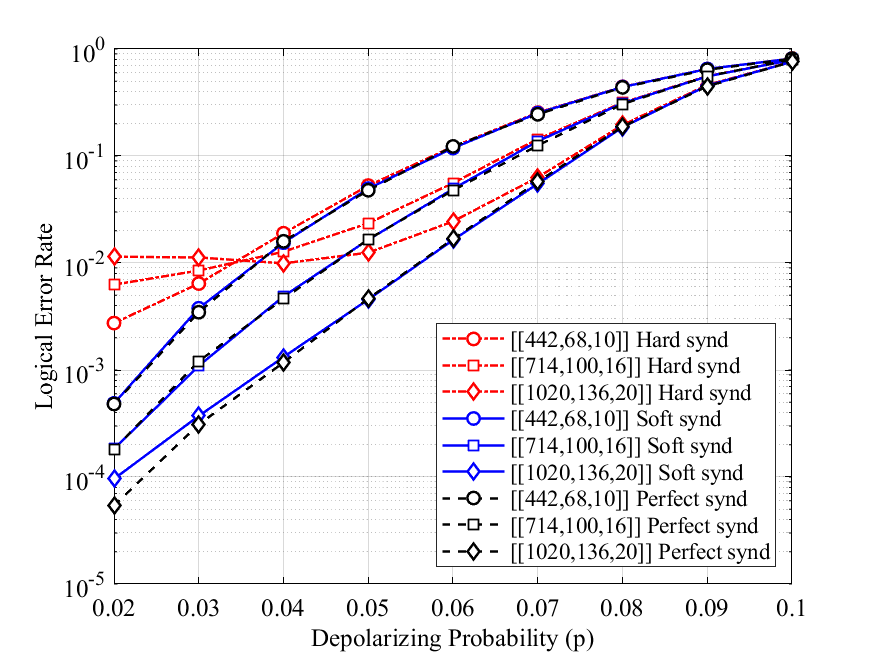}
    \caption{Two-threshold phenomenon observed for `Hard Syndrome' (which uses standard MSA), but the `Soft Syndrome' setting (which uses modified MSA) is devoid of this issue.}
    \label{fig:two_thresholds}
    \vspace{-10pt}
\end{figure}

As one might expect, the threshold for the `Hard Syndrome' case is worse than `Perfect Syndrome', since the decoder is unchanged.
But, more interestingly, for the `Hard Syndrome', when we fix the variance of the Gaussian noise affecting the syndrome and sweep the qubit depolarizing rate, below what we identify as the usual decoding threshold, there arises a ``second (inverted) threshold''; below this (lower) threshold, the logical error rate creeps up again because the syndrome error dominates the qubit error much more strongly (see Fig.~\ref{fig:two_thresholds}).
This suggests that, under realistic settings, it is insufficient to only optimize the threshold, and one needs to avoid such problematic behavior below the threshold.

Next, we propose the modified MSA decoder with justification for our new rules for updating nodes as well as the syndrome LLR. 
We show that the resulting decoder does not suffer from this spurious ``second threshold'' phenomenon.
Besides, the decoding threshold is now almost as good as that for the `Perfect Syndrome' setting.
Hence, our work paves the way for efficient and effective decoding of good QLDPC codes under a physically motivated phenomenological noise model, without the use of repetitive syndrome measurement rounds.

The paper is organized as follows. In Section~\ref{Sec:Prelims} we introduce the necessary notations and explain the syndrome-based MSA and its variants. In Section~\ref{Sec:SoftSyndromeDecoders} we motivate the soft syndrome decoders and explain our modification to correct the noisy syndromes. Section~\ref{Sec:SimResults} presents simulation results showing the benefits of our modified MSA.
Finally, Section~\ref{Sec:Conclusions} concludes this paper with discussion on future work.
\section{Preliminaries}
\label{Sec:Prelims}
\subsection{Pauli Operators and CSS Codes}

The single-qubit Pauli operators are ${\rm I, X, Z, Y} = \imath {\rm X Z}$ and the $n$-qubit Pauli group $\mathcal{P}_n$ consists of tensor products of single-qubit Paulis up to a quaternary global phase $1,\imath, -1,$ or $-\imath$, where $\imath = \sqrt{-1}$.
There is a useful binary representation $[a,b] \in \{0,1\}^{2n}$ of Pauli operators, e.g., ${\rm X} \otimes {\rm Z} \otimes {\rm Y} \mapsto [a,b] = [1 0 1, 0 1 1]$, so that entries of $a$ (resp. $b$) indicate which qubits are acted upon by ${\rm X}$ (resp. ${\rm Z}$), and $a_i=b_i=1$ (resp. $0$) indicates a ${\rm Y}$ (resp. ${\rm I}$).
An $[[n,k,d]]$ CSS code~\cite{calderbank1996quantum_exists} encodes $k$ logical qubits into $n$ physical qubits, and $d$ is the minimum distance of the code that indicates the minimum number of qubits on which the error channel must act to disturb the logical information. 
The CSS code is defined by $(n-k)$ commuting Pauli operators, called \emph{stabilizers}, some of which are purely ${\rm X}$-type, $[a,0]$, and the others are purely ${\rm Z}$-type, $[0,b]$. 
Hence, the stabilizers can be represented using two binary parity-check matrices $H_{\mathrm{X}}$ (whose rows map to $[a,0]$) and $H_{\mathrm{Z}}$ (whose rows map to $[0,b]$) that must satisfy $H_{\mathrm{X}} H_{\mathrm{Z}}^\text{T} = 0$ so that these Pauli operators commute. 

\subsection{Noise Model}

The single-qubit depolarizing channel, also known as the memoryless Pauli channel, is a widely studied error model characterized by the depolarizing probability $p$: the channel randomly introduces a Pauli error 
according to the probabilities $P_{\rm X}= P_{\rm Y}=P_{\rm Z}= p/3$ and $P_{\rm I}= 1-p$. 
For the physical data qubits, we assume an i.i.d. depolarizing error model, i.e., each physical qubit is affected by an independent depolarizing error.
Let $\mathbf{e} = [ \mathbf{e_{\mathrm{X}}}, \mathbf{e_{\mathrm{Z}}} ]$ be the binary representation of a Pauli error acting on the $n$ qubits. The corresponding syndrome is computed as $[\mathbf{s_{\mathrm{X}}},~\mathbf{s_{\mathrm{Z}}}] = [H_{\mathrm{Z}} \, \mathbf{e_{\mathrm{X}}}^{\text{T}}\, , \, H_{\mathrm{X}} \, \mathbf{e_{\mathrm{Z}}}^{\text{T}} ]^{\text{T}} \, (\bmod\ 2).$ Since we consider decoding of $H_{\mathrm{X}}$ and $H_{\mathrm{Z}}$ separately, we avoid using subscripts $\mathrm{X}$ and $\mathrm{Z}$ in syndrome $\mathbf{s}$.

\subsection{Syndrome Noise Model}

As mentioned in Section~\ref{Sec:Intro}, we model a realistic syndrome measurement by a perfect measurement with an ideal bipolar ($\pm 1$) outcome followed by i.i.d. symmetric Gaussian noise on each component of the syndrome.
After 
introducing symmetric Gaussian noise $n_i \sim \mathcal{N}(0,\sigma^2)$ 
to the ideal syndrome (component) $s_i \in \{ \pm 1\}$, we have a noisy soft syndrome $\tilde{r_i} = s_i + n_i$ which when conditioned on $s_i$ is distributed normally: $[ \tilde{r_i} \mid s_i = 1 ] \sim N(+1, \sigma^2)$ and $[ \tilde{r_i} \mid s_i = -1 ] \sim N(-1, \sigma^2)$. Other relevant asymmetric syndrome measurement models are discussed in \cite[Section 1.4]{ImprovedDecodingSoftInfo_2021}. 

Based on the normally distributed soft syndromes, we can compute the syndrome log-likelihood ratios (LLRs) as 
\begin{align}\label{eq:gamma_i}
    \gamma_i \coloneqq \log\frac{\Pr(\tilde{r_i} | s_i = +1)}  {\Pr(\tilde{r_i}| s_i = -1)} = \frac{2 \tilde{r_i}}{\sigma^2}.
\end{align} 
Since the noise is zero-mean, the sign of the soft syndrome, $\text{sgn}(\gamma_i)$, indicates if the parity-check $i$ is satisfied ($\text{sgn}(\gamma_i) = +1$) or unsatisfied ($\text{sgn}(\gamma_i) = -1$).
This is exactly same as the measured syndrome $s_i$ used in conventional min-sum decoder. 
However, its magnitude, $|\gamma_i|$, provides valuable soft information about the reliablility of the measured syndrome.

\subsection{QLDPC Codes and Syndrome Min-Sum Algorithm (MSA)}
\label{subsec:SyndromeDecoders}
\begin{figure}     
\centering
\begin{subfigure}[b]{0.17\textwidth} 
\centering 
\includegraphics[width=\textwidth]{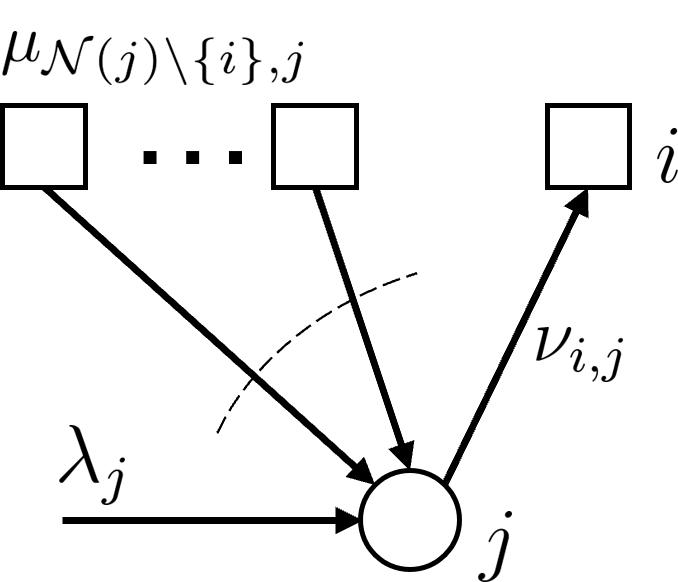} 
\caption{Variable node update} 
\label{fig:VNU} 
\end{subfigure}
\begin{subfigure}[b]{0.01\textwidth}
\hspace{15pt}
\end{subfigure}
\begin{subfigure}[b]{0.18\textwidth} 
\centering 
\includegraphics[width=\textwidth]{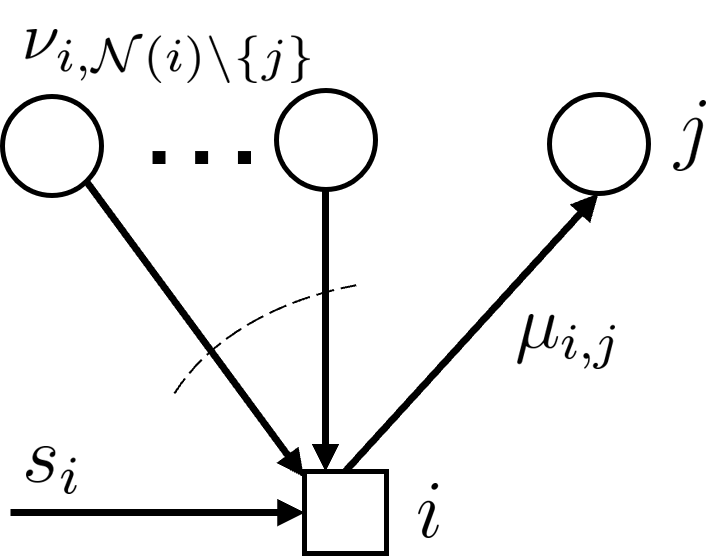} 
\caption{Check node update} 
\label{fig:CNU} 
\end{subfigure}
\caption{Node updates for MSA based on Eqs. \eqref{eq:VNMSA} and \eqref{eq:CNMSA}.
}
\label{fig:UpdateRules}
\end{figure}

A CSS-QLDPC code has sparse stabilizer generators and is represented by a pair of bipartite (Tanner) graphs whose biadjacency matrices are the stabilizer matrices $H_{\mathrm{X}}$  and $H_{\mathrm{Z}}.$ The Tanner graphs have $n$ variable nodes, denoted by $j \in \{1,2,\ldots, n\}$, represented by circles in Fig. \ref{fig:UpdateRules}, and $m$ check nodes, denoted by $i \in \{1,2,\dots, m\}$, represented by squares in Fig. \ref{fig:UpdateRules}. 
If the $(i,j)$-th entry of the binary parity-check matrix is not zero, then there is an edge between variable node $i$ and check node $j.$
The sets of neighbors of variable node $j$ and check node $i$ are denoted by $\mathcal{N}(j)$ and $\mathcal{N}(i)$, respectively. 

Like classical LDPC codes, QLDPC codes also employ decoders based on iterative 
BP or message-passing algorithms. 
The fundamental difference between the decoders of QLDPC codes and LDPC codes is that, the decoders for the former take as input the syndrome obtained after the stabilizer measurements, whereas decoders for the latter take as input a noisy version of the codeword. 
The QLDPC decoder's goal is to output an error pattern whose syndrome matches the measured syndrome. Let us first consider the case when these stabilizer measurements are perfect, meaning there is no noise in the syndrome measurement process. 

Suppose that the coded qubits are corrupted by an ${\rm X}$-type or ${\rm Z}$-type error, corresponding to the binary error vector $\mathbf{e} = [e_1, e_2, \cdots, e_n]$ whose entries are realizations of independent Bernoulli random variables with parameter $p$.   
Given binary syndrome $\mathbf{s}$ vector of length $m$, the decoder performs a finite number of message-passing iterations over the Tanner graph to compute \emph{a posterior probabilities} $\Pr(e_j| \mathbf{s})$, for $j \in \{1,\ldots,n\}$, corresponding to error bit $e_j$ conditioned on the measured syndrome $\mathbf{s}$. 

Next, we briefly describe the min-sum algorithm (MSA) based decoder. For this purpose, we first introduce the required notations.
Denote by $\mu_{i,j}$ the message sent from check node $i$ to variable node $j$. Similarly, the message sent from variable node $j$ to check node $i$ is denoted by $\nu_{i,j}$.
The measured syndrome value is $s_i \in \{ \pm 1 \}$, $i \in \{1,2,\ldots,m\}$.
Since the error $e_j$,  $j \in \{1,2, \ldots n\}$ , at variable node $j$ is a realization of Bernoulli random variable with parameter $p,$ the log-likelihood ratio (LLR), denoted by $\lambda_j,$ corresponding to the $j$th variable node is given by $\lambda_j = \ln \left(\frac{1-p}{p}\right).$

In the first decoding iteration, all the outgoing messages from variable node $j$, for all $j$, are initialized to $\lambda_j.$ In subsequent iterations, for $i \in \mathcal{N}(j),$ message $\nu_{i,j}$ 
is computed using the update rule
\begin{equation}
\label{eq:VNMSA}
\nu_{i, j} = \lambda_j + \sum_{i' \in \mathcal{N}(j) \backslash \{ i \} }\mu_{i',j}.    
\end{equation}

For $j \in \mathcal{N}(i),$ the message $\mu_{i,j}$ 
is computed using the rule
\begin{equation}
    \label{eq:CNMSA}
\mu_{i, j} = \left( s_i \cdot \prod_{j' \in \mathcal{N}(i) \backslash \{ j \}}\textrm{sgn}(\nu_{i,j'}) \right) \cdot \left(\underset{ j'\in \mathcal{N}(i)\backslash \{ j \}}{\min}\left | \nu_{i,j'} \right |\right).
\end{equation}
 The sign function is defined as $\text{sgn}(A) = -1$ if $A<0$, and $+1$ otherwise. The first and second parenthesis in Eq. \eqref{eq:CNMSA} gives the sign and the magnitude of the computed check-to-variable message $\mu_{i,j}$, respectively. Iteration will be indicated as a superscript of $\mu_{i,j}$ and $\nu_{i,j}$ when required. Min-sum update rules are less complex than 
 that of BP decoders (which are never used in practice), but 
 as the min-sum approximation overshoots the BP messages,  
 the MSA is typically modified using a normalized MSA \cite{05CDEFH}, wherein check node messages are multiplied (in Eq. \eqref{eq:CNMSA}) by a scalar $\beta$ as a correction factor, where $0 < \beta < 1$.

A min-sum decoder employs the update functions in Eq.~\eqref{eq:VNMSA} and Eq.~\eqref{eq:CNMSA} in each iteration $\ell \le \ell_{\max}$ to determine the error at each variable node.
The error estimate $\hat{x}_j$ at the variable node $j$ is computed using
\begin{equation}
\label{eq:err_est}
    \hat{x}_j = \hat{\Phi}(\lambda_j, \mu_{i,j}) = \mathrm{sgn}\left(\lambda_j + \sum_{i \in \mathcal{N}(j) }\mu_{i,j} \right),
\end{equation}
for all $j \in \{1,2,\ldots,n\}$, i.e., 
it is based on the sign of the \emph{decision update function} output, $\hat{\Phi}(\lambda_j, \mu_{i,j})$, that uses the LLR value and \emph{all} incoming check node messages to the variable node $j$, where $i \in \mathcal{N}(j)$. 

\textbf{Halting Criterion: } 
The output of the decoder at the $\ell$-th iteration, denoted by $\mathbf{\hat{x}}^{(\ell)}=(\hat{x}^{(\ell)}_1,\hat{x}^{(\ell)}_2,\ldots,\hat{x}^{(\ell)}_n)$, is used to check whether all parity-check equations are matched, i.e., whether the syndrome at $\ell^{\text{th}}$ iteration ${\hat{\mathbf{s}}^{(\ell)}} \coloneqq  {\bf \hat{x}^{(\ell)}}\cdot {\bf H}^{\rm T}$ is equal to the measured syndrome ${\bf s}$, in which case iterative decoding terminates and outputs $\bf \hat{x}^{(\ell)}$ as the error vector.
Otherwise, the iterative decoding steps continue until a predefined maximum number of iterations, denoted by $\ell_{\max}$, is reached. The decoding is deemed successful if the true error pattern is found, i.e., $\mathbf{\hat{x}}^{(\ell)} = \mathbf{e}$. Otherwise, the decoding is said to have failed. 
Miss-correction as referred to in the literature of classical coding theory occurs when post-correction step with the estimated error pattern results in a logical error. This is also classified as an error correction failure.

Note that the MSA decoder is exactly the ``Perfect Syndrome'' decoder, described in Section~\ref{Sec:Intro}, when the syndrome measurement process is noiseless.
When we have a noisy observation of the syndrome, recall that an estimate of the syndrome is obtained by passing the observations to a thresholding function. This estimated syndrome is fed to the MSA decoder to get the `Hard Syndrome' decoder defined earlier. 
 \section{Soft Syndrome Decoder}
\label{Sec:SoftSyndromeDecoders}

As discussed above, under noisy syndrome, the MSA decoder does not use the soft information available in the measured syndrome and treats the thresholded syndrome as the correct syndrome. 
Due to this reason, the decoder typically obtains error pattern estimates only close to the true error pattern. 
Also, note that the halting condition is never met in this case, so the decoder runs until the maximum number of iterations and declares a decoding failure.
In particular, using a noisy syndrome has two disadvantages:  it makes the iterative decoder's halting condition ambiguous, even after the decoder converges to the true error estimate, and drives the decoder to produce error estimates matching the wrong syndrome.
In this section, we propose a modified MSA decoder that uses the soft information available in the measured syndrome to address these issues. As mentioned in Section~\ref{Sec:Intro}, we refer to the proposed decoder as `Soft Syndrome' (MSA).

To use the soft information from the observed syndrome, we modify the Tanner graph by adding another type of nodes, called ``virtual variable nodes'', in addition to the existing variable and check nodes. A virtual variable node is added corresponding to each of the $m$ check nodes. There is an edge between a virtual variable node and the corresponding check node. The connections between the variable and check nodes are not altered. Hence, the standard MSA decoder still holds. 

The initialization and update rules at the variable nodes for the `Soft Syndrome' decoder is same as that of the MSA decoder described in Section~\ref{Sec:Prelims}.
We now describe the update rules at the check nodes and virtual variable nodes for the `Soft Syndrome' decoder. 
In addition to the measured syndrome $\mathbf{s}$, a binary vector of length $m$, the `Soft Syndrome' decoder receives the respective reliability (i.e., syndrome LLR) values $\gamma_i$ as defined in Eq.~\eqref{eq:gamma_i}. The goal of this `Soft Syndrome' iterative decoder in the noisy syndrome setting is to find an error pattern that matches the \emph{potentially revised} syndrome $\tilde{\mathbf{s}}$ using the syndrome reliability $\gamma$ and the qubit LLR $\lambda$. 


The magnitude of the soft syndrome LLR value determines the reliability of the check information. To deploy the `Soft Syndrome' decoder, we determine a \emph{cutoff} above which the soft syndrome value is deemed reliable and below which it is not. If the syndrome is reliable, i.e., if the $|\gamma_i|$ corresponding to check node $i$ is greater than the predefined cutoff, then we use the conventional check node update rule given in Eq.~\eqref{eq:CNMSA}. Choosing this cutoff value denoted by $\Gamma$ is important and it can be optimized to improve the decoding performance. 
There are two main changes to the minimum rule used at the check nodes. With the soft syndrome $\gamma_i$ available from the virtual variable node, first, we compute the check to variable message as the minimum of magnitude of the extrinsic variable node messages as well as the soft syndrome magnitude $|\gamma_i|$. 
For $ j\in \mathcal{N}(i)$, the magnitude and sign of message $\mu_{i,j}$ are given by  
\begin{align}
    \label{eq:mod_CUD_mag}
    |\mu_{i,j}| &= 
    \begin{cases}
     \left(\underset{ j'\in \mathcal{N}(i)\backslash \{ j \}}{\min}\left | \nu_{i,j'} \right |\right) & 
     \text{for } |\gamma_i| > 
     \Gamma, \\ \\
      \left(\underset{ j'\in \mathcal{N}(i)\backslash \{ j \}}{\min}(\left | \nu_{i,j'} \right |,|\gamma_i|)\right) & \text{otherwise;}
      \end{cases} \\
     \label{eq:mod_CUD_sign}
    \mathrm{sgn}(\mu_{i,j}) &= 
    \left( s_i \cdot \prod_{j' \in \mathcal{N}(i) \backslash \{ j \}}\textrm{sgn}(\nu_{i,j'}) \right).    
\end{align}

Second, we update the soft syndrome, both magnitude and reliability, based on the magnitude and sign of all the incoming messages to the check node from its neighboring variable nodes, thus effectively pushing the estimated syndrome towards the correct syndrome. 
In the $\ell$-th iteration, the estimated sign and reliability of syndrome corresponding to check node $i$ is denoted by $\tilde{s}_i^{(\ell)}$ and $\tilde{\gamma}_i^{(\ell)}$, respectively. 
In the first iteration, we initialize $\tilde{s}_i^{(1)} = s_i$ and $\tilde{\gamma}_i^{(1)} = |\gamma_i|.$ 
In the subsequent iterations, the reliability $\tilde{\gamma}_i^{(\ell)}$ of the estimated syndrome corresponding to the check node $i$ is updated using
\begin{equation}
    \label{eq:check-to-vv}
    \tilde{\gamma}_i^{(\ell)} =
    \underset{ j'\in \mathcal{N}(i)}{\min}\left | \nu_{i,j'} \right |
\end{equation}
if  $\underset{j' \in \mathcal{N}(i)}{\min} \left|\nu_{i,j'}\right| > \left|\tilde{\gamma}_i^{(\ell-1)}\right|$ and $\tilde{s}_i^{(\ell-1)} = \prod_{j' \in{\mathcal{N}(i)}} \mathrm{sgn}(\mu_{i,j'})$, otherwise $\tilde{\gamma}^{(\ell)}_i = \tilde{\gamma}_i^{(\ell-1)}.$ 

The sign of the estimated syndrome corresponding to the check node $i$ is updated using 
\begin{equation}
    \label{eq:check-to-vv-sign}
    \tilde{s}_i^{(\ell)} \mapsto (-1) \cdot \tilde{s}_i^{(\ell-1)}
\end{equation}
if  $\underset{j' \in \mathcal{N}(i)}{\min} \left|\nu_{i,j'}\right| > \left|\tilde{\gamma}_i^{(\ell-1)}\right|$ and $\tilde{s}_i^{(\ell-1)} \neq \prod_{j' \in{\mathcal{N}(i)}} \mathrm{sgn}(\mu_{i,j'})$, otherwise $\tilde{s}_i^{(\ell)} =  \tilde{s}_i^{(l-1)}.$

\begin{figure*}[t]
    \centering
    \includegraphics[width=\textwidth]{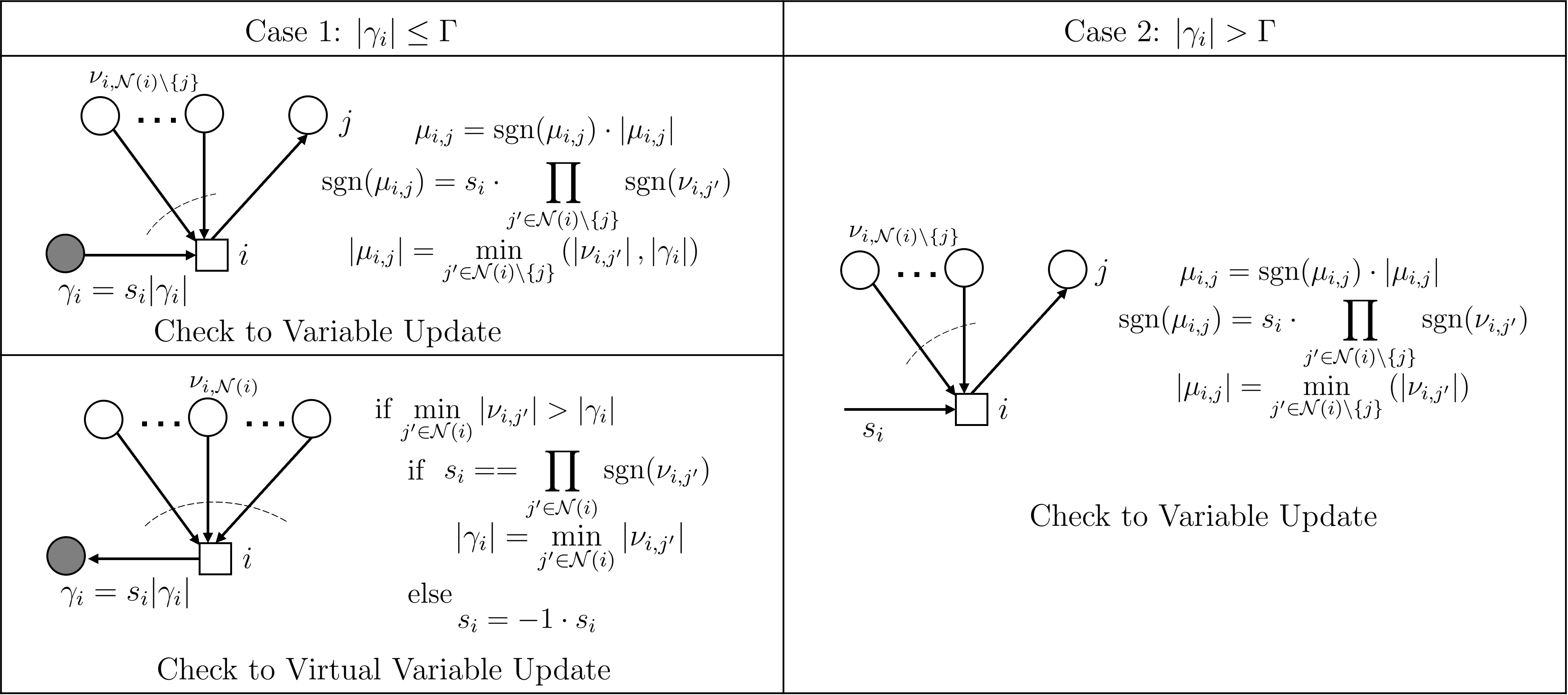}
    \caption{Modification of the check node update for the `Soft Syndrome' decoder based on a predefined cutoff, $\Gamma$. In addition to the `minimum' update, check to virtual variable updates (shaded variable node) are updated conditionally.}
    \label{fig:SoftSyndromeDecoder}
    \vspace*{-0.35cm}
\end{figure*}

The halting criterion for the `Soft Syndrome' decoder is whether all parity-check equations are matched to these updated signs $\tilde{s}_i^{(\ell)} ~\forall i \in \{1,2,\ldots,m\}$, i.e., whether the syndrome at $\ell^{\text{th}}$ iteration, ${\bf \hat{x}^{(\ell)}}\cdot {\bf H}^{\rm T}$ is equal to the $\tilde{\mathbf{s}}$, in which case decoding terminates and outputs $\bf \hat{x}^{(\ell)}$ as the error vector. The modified decoding rules are illustrated in Fig.~\ref{fig:SoftSyndromeDecoder}.

Though in the simulation results in Section \ref{Sec:SimResults}, we use the `Soft Syndrome' discussed above, we also identify another variant where one do not update the reliability of the estimated syndrome given in Eq.~\eqref{eq:check-to-vv}, and only update the sign of the estimated syndrome. Upon selecting a suitable cutoff, this variant of `Soft Syndrome' decoder also can correct both data and syndrome errors. We will analyze performance of these variants as well the effect of the cutoff in future work. 

Note that we are not using an explicit syndrome code for protecting the syndrome bits as in~\cite{ashikhmin_data_syndrome_codes, kuo2021DataSyndromeBPdecoding}, instead relying on the message passing algorithm to infer what the syndromes should be. Using this modified MSA decoder, we can avoid the overhead of repeated measurements as we can identify the measurement errors in the instances where this decoder converges. 

\section{Simulation Results}
\label{Sec:SimResults}
\subsection{QLDPC Codes and Simulation Setup}
We use \emph{lifted product} (LP) QLDPC codes proposed by Panteleev and Kalachev \cite{panteleev2020quantumLinearMinD} which provide flexibility in constructing finite length QLDPC codes from good classical (and quantum) \emph{quasi-cyclic (QC)} \cite{Fossorier04} LDPC codes. We choose QC-LDPC codes as constituent classical LDPC codes to construct LP code families for demonstrating our simulation results. For example, from the $[155, 64, 20]$ Tanner code \cite{tannercode} with quasi-cyclic base matrix of size $3 \times 5$ and circulant size $L=31$, we obtain the $[[1054, 140, 20]]$ \emph{LP Tanner code} \cite{nithin_GKP_QLDPC}. To show the decoding threshold plots, we also chose the LP code family described in \cite[Table II]{nithin_GKP_QLDPC}. LP-QLDPC codes with increasing code length and minimum distance ($d = 10, 16, 20, 24$) are chosen to demonstrate the thresholds.  
The syndrome-based MSA is chosen with a normalization factor of $\beta$ set to $0.75$ empirically, and has a preset maximum of $\ell_{\max} = 100$ iterations. For the Monte-Carlo simulations, specifically for the threshold plots, we collect at least 10,000 logical errors (sufficient to avoid statistical errors). The Cutoff $\Gamma$ is set to $10$ for Fig.~\ref{fig:two_thresholds} and to $5$ for the rest of the simulation plots. 

Different scenarios of simulation setups --- `Perfect', `Hard', and `Soft' syndrome decoding as described in Section~\ref{Sec:Intro} ---
are considered, with depolarizing noise as the underlying noise for the physical qubits. 
We also denote the respective MSA decoders as `perfect synd', `hard synd', and `soft synd'.

\subsection{Threshold Plots}

\begin{figure*}
\centering
\vspace*{-0.5cm}
\begin{subfigure}[t]{0.45\linewidth}
    \centering
    \includegraphics[trim=10 0 10 0,clip, width=\textwidth]{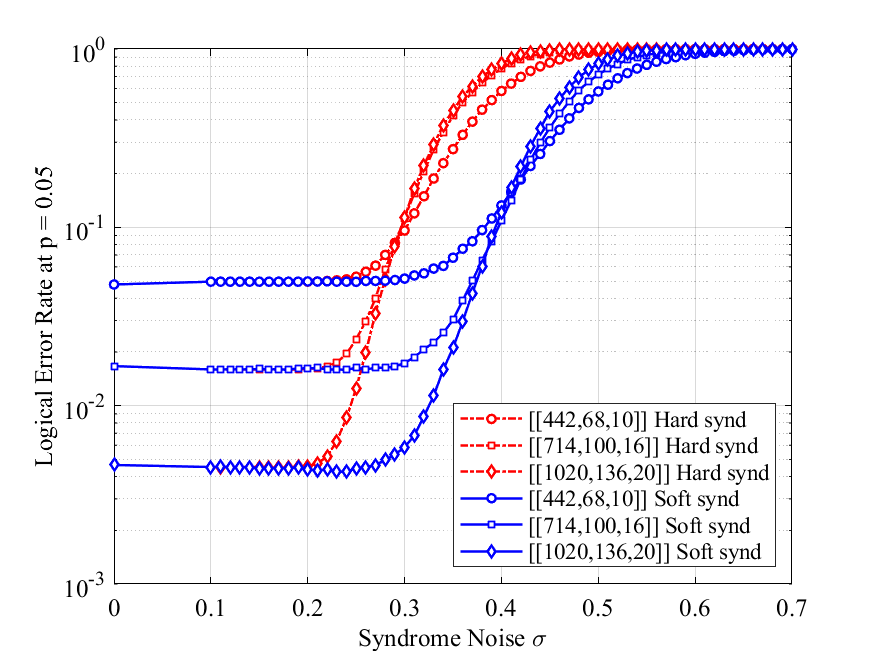}
    %
    \caption{Logical Error Rate for a fixed depolarizing probability of $p = 0.05$ is plotted against the syndrome noise standard deviation $\sigma$ for the QLDPC code family with increasing code length and distance. We can observe the increase in $\sigma$ threshold from $\approx 0.25$ for `Hard Syndrome' to $\approx 0.4$ for `Soft Syndrome'.}
    \label{fig:SyndNoiseThreshold}
\end{subfigure} \qquad    
\begin{subfigure}[t]{0.45\linewidth}
    \centering
    \includegraphics[trim=10 0 10 0,clip, width=\textwidth]{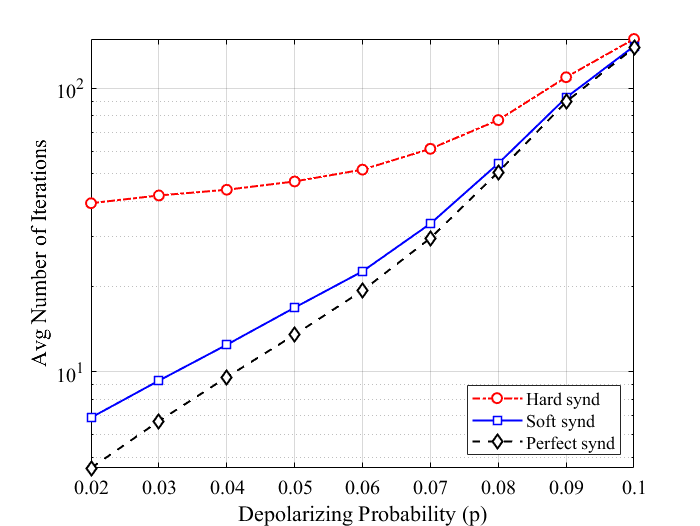}
    \caption{For LP Tanner code with $\sigma = 0.3$, average number of iterations while considering decoding over both $H_{\text{X}}$ and $H_{\text{Z}}$ matrices shows that `Soft Syndrome' decoder requires far less decoding iterations to correct both data and syndrome errors compared to the `Hard Syndrome' decoder.}
    \label{fig:AvgDecoderIters}
\end{subfigure}
\caption{Benefits of the `Soft Syndrome' decoder: (a) syndrome noise threshold, (b) average number of iterations.}
\end{figure*}

\textbf{Threshold on Syndrome Noise:} 
To observe the syndrome noise threshold, we plot in Fig. \ref{fig:SyndNoiseThreshold} the logical error rate for a fixed depolarizing probability of $p = 0.05$ against different values of the syndrome noise standard deviation $\sigma$. 
In this setting, the `Hard Syndrome' decoder 
produces an early threshold at syndrome noise $\sigma \approx 0.25$. Above this syndrome noise level, a hard decoder cannot improve the decoding performance even by increasing the code length arbitrarily. However, with the `Soft Syndrome' decoder we see that the threshold point is at a higher syndrome noise value of $\sigma \approx 0.4$. This indicates that the modified MSA decoder is more robust and can tolerate noisier syndromes. 

\begin{figure*}     
\centering
\vspace*{-0.5cm}
\begin{subfigure}[t]{0.45\linewidth} 
\centering 
\includegraphics[trim=10 0 10 0,clip, width=\textwidth]{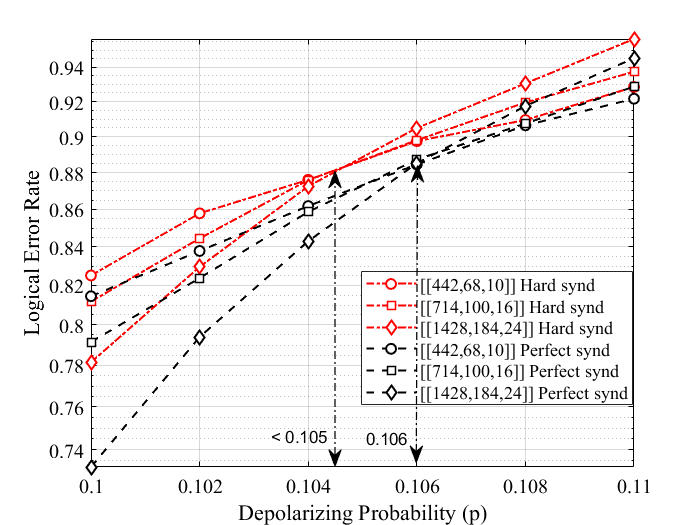}
\caption{Thresholds for the LP-QLDPC codes from LP118 code family in \cite{nithin_GKP_QLDPC}: `Perfect Syndrome' vs `Hard Syndrome'} \label{fig:LP118_PerfectNoisyWithoutSoftSynd} 
\end{subfigure} \qquad
\begin{subfigure}[t]{0.45\linewidth} 
\centering 
\includegraphics[trim=10 0 10 0,clip, width=\textwidth]{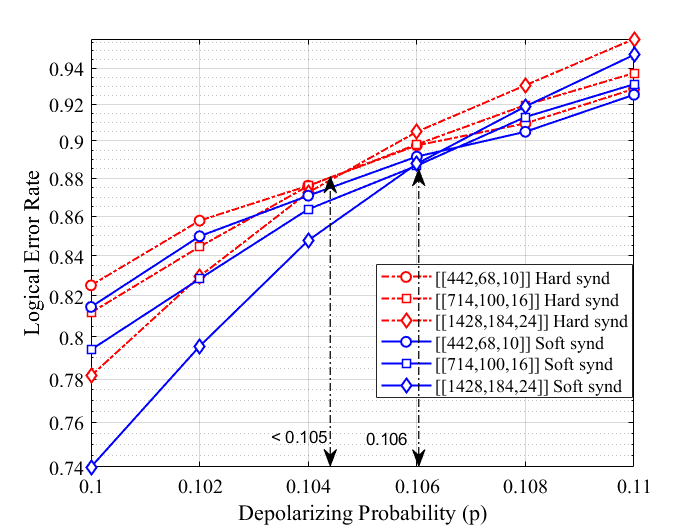} 
\caption{Thresholds for the LP-QLDPC codes from LP118 code family in \cite{nithin_GKP_QLDPC}: `Soft Syndrome' vs `Hard Syndrome'} \label{fig:LP118_WithWithoutSoftSynd} \end{subfigure}
\caption{The set of curves in figures (a) and (b) correspond to use of different decoders in the perfect and noisy syndrome setting. The transition of the curves with increasing $p$ signifies an error threshold.}
\label{fig:threshold_dv3dc_5}
\vspace*{-0.35cm}
\end{figure*}

This experiment also gives us the range of syndrome noise where `Soft Syndrome' decoder performs close to the `Perfect Syndrome' decoder at $p=0.05$ (from Fig.~\ref{fig:two_thresholds}).  
Interestingly, at low levels of syndrome noise ($\sigma < 0.25$), the `Soft Syndrome' decoder is able to perform at least as good as 
the `Perfect Syndrome' decoder. This is a particularly encouraging noise regime where the soft information about noisy syndromes can be dealt with the modified MSA intrinsically, without repeating measurements or requiring single-shot property in the code.
For our subsequent simulations, we use the syndrome noise $\sigma = 0.3$ to observe the decoding threshold and the logical error rate suppression with respect to qubit depolarizing probability.

\vspace{5pt}

\textbf{Threshold on Depolarizing Noise: } 
Using the same QLDPC code family and a fixed syndrome noise standard deviation $\sigma = 0.3$, we plot the logical error rate against depolarizing probability for the decoders of interest in Fig.~\ref{fig:threshold_dv3dc_5}. 
One can see the transition from error suppression to error enhancement with increasing $p$ which signifies the existence of the error threshold. 
As expected, since we use the same standard MSA decoder, we see in Fig.~\ref{fig:LP118_PerfectNoisyWithoutSoftSynd} that the threshold for the `Hard Syndrome' case is worse than `Perfect Syndrome'.
More interestingly, when we look at lower $p$ in Fig.~\ref{fig:two_thresholds}, below what we identify as the usual decoding threshold, there exists another crossover point for the `Hard Syndrome' case where the logical error rate rises up again because the syndrome error starts to dominate the qubit error much more strongly. 
However, the `Soft Syndrome' curves in both Fig.~\ref{fig:two_thresholds} and Fig.~\ref{fig:LP118_WithWithoutSoftSynd} show that, besides achieving the same decoding threshold as `Perfect Syndrome', the modified MSA is also devoid of the spurious ``second threshold'' phenomenon.

\subsection{Faster Decoder Convergence}

As we discussed in Section~\ref{Sec:Intro} and Section~\ref{Sec:SoftSyndromeDecoders}, the stopping criterion of an iterative decoder in the presence of measurement errors is ambiguous. 
To understand the correct decoding performance of these decoders in the presence of syndrome errors, we do not automatically declare a logical error if the decoder exhausts the maximum number of iterations and fails to find an error matching the (evolved) syndrome.
Instead, we declare a logical error if the true error `plus' the error estimate produced by the decoder has a non-trivial syndrome or results in a logical error. This gives a fair chance to the iterative decoders in the presence of syndrome errors. 

Even in the presence of noisy syndrome, the `Soft Syndrome' decoder is able to correct the syndrome and data errors quickly to converge to the true error pattern. The faster convergence of `Soft Syndrome' decoder is shown in Fig.~\ref{fig:AvgDecoderIters}, which clearly shows that it requires an average number of decoding iterations very close to that of the `Perfect Syndrome' decoder. 
Hence, under a reasonable amount of syndrome noise, besides avoiding repeated syndrome measurements and single-shot property, the modified MSA also does not suffer from a penalty in decoding latency.
These are very attractive features of the decoding procedure for practical implementations.

\section{Conclusions and Future Directions}
\label{Sec:Conclusions}
In the past few months it has been proven that there is no fundamental obstacle to construct QLDPC code families with optimal parameters.
However, one still needs to develop fast and effective decoders for these codes under practical settings where the syndrome measurement process can itself be noisy.
In this work, we considered the min-sum algorithm (MSA) based iterative message passing decoder for QLDPC codes.
We showed that standard MSA is affected severely by syndrome noise, and we proposed a modified MSA that is robust to a reasonable amount of syndrome noise even without repeated measurements.
In future work, we will further investigate systematic ways to analyze iterative decoding under noisy syndromes, and also consider realistic circuit-level noise.

\IEEEtriggeratref{12}
\bibliographystyle{IEEEtran}
\bibliography{refs}

\end{document}